\documentclass[a4paper]{jpconf}
\usepackage{graphicx}
\usepackage[vcentermath]{youngtab}
\usepackage{setspace,amsmath,bm,amssymb} 
\begin{document}
\title{Magnetoplasmons and SU(4) symmetry in graphene}

\author{Andrea M Fischer$^1$, Rudolf A R\"omer$^1$ and Alexander B
Dzyubenko$^{2,3}$}

\address{$^1$ Department of Physics and Centre for Scientific Computing,
University of Warwick, Coventry, CV4 7AL, UK}
\address{$^2$ Department of Physics, California State University -  Bakersfield,
Bakersfield, California 93311, USA}
\address{$^3$ General Physics Institute, Russian Academy of Sciences, Moscow
119991, Russia
}

\ead{A.M.Fischer@warwick.ac.uk}

\begin{abstract}
We study magnetoplasmons or neutral collective
excitations of graphene in a
strong perpendicular magnetic field, which can be modelled as bound electron-hole
pairs. The SU(4) symmetry of graphene arising from spin and valley pseudospin
degrees of freedom is explored using Young diagrams to correctly predict the
degeneracies of these excitations. The multiplet structure of the states is
identical to that of mesons composed of first and second generation quarks.
\end{abstract}

\section{Introduction}
Graphene is a single layer of graphite or set of carbon atoms $sp^2$ bonded
together to form a honeycomb lattice. The electron in the remaining $p_z$
orbital is delocalised and responsible for graphene being an excellent
electrical conductor \cite{Per10}. Graphene is associated with numerous other
superlatives such as strongest material \cite{LeeWKH08} and of course thinnest
material; it is
the first example of an atomically thick membrane. Such properties make it an
excellent candidate for industrial applications, although in many cases there
are obstacles to be overcome before this can be put into practice \cite{Gei09}.
Aside from
its potential usefulness, graphene is scientifically interesting in its own
right. Unlike other materials, including the 2D electron gas (2DEG), electrons
in graphene have a linear
dispersion relation close to the Fermi energy and in this sense behave
quasirelativistically \cite{CasGPN09}. Effects such as Klein
tunnelling, which were previously the dominion of high-energy particle physics, have now
been observed in graphene \cite{Bee08}. For an undoped system, the Fermi energy
is located at
zero energy. The Fermi surface consists of six points which coincide with the
six corners of the hexagonal Brillouin zone. Only two of these points are
inequivalent and are commonly termed the $\mathbf{K}$ and $\mathbf{K}'$ valleys.
They may also be referred to as pseudospins, since they result in the single
particle wavefunctions having a spinor form \cite{Per10}.

In these proceedings we examine a monolayer of graphene in the presence of a
strong perpendicular magnetic field of magnitude $B$. The Landau level (LL) energies for
graphene are, ${\epsilon_n = {\rm sign}(n) \hbar v_\mathrm{F}
\sqrt{2|n|}/\ell_B}$, where $v_\mathrm{F}\sim 10^6 \mathrm{ms^{-1}}$ is the
Fermi velocity, $\ell_B=\sqrt{\hbar c/eB}$ is the magnetic length and
$\mathrm{sign}(0)=0$. There is also the usual LL degeneracy resulting from
the different possible guiding centres for electronic cyclotron orbits. This is
described by the oscillator quantum number, $m$, in the
symmetric gauge. We consider magnetoplasmons (MPs), which are essentially excitons created from a ground
state described by the filling factor $\nu=2$, where all states with LL indices
$n \le 0$ are occupied and all other states are empty. An exciton is formed when
one of the electrons in this ground state is excited to a state in a higher
lying LL and becomes bound to the hole it leaves behind. We focus on excitons
with the lowest possible excitation energies, namely those with the hole in the
$n=0$ LL and the electron in the $n=1$ LL. Taking into account the
spin/pseudospin states of the electron and hole, this yields 16 resonant
excitations. These are collective excitations on account of the LL degeneracy
and also due to the fact that excitons with different spin/pseudospin characters
may be mixed by the two-body Coulomb interaction. In the following, we study the
SU(4) symmetry in graphene, which arises from the two spin and two pseudospin
degrees of freedom, denoted by $\uparrow, \downarrow$ and $\Uparrow,
\Downarrow$ respectively. This leads to a direct analogy
between MPs in graphene and mesons composed of either first or second
generation quarks. We then review the dispersion relation for MPs in
pristine graphene and finally examine the states which become localised in the
presence of a single Coulomb impurity. In both cases we observe the
degeneracies predicted by the SU(4) symmetry and Young diagram techniques.
%%%%%%%%%%%%%%%%%%%%%%%%%%%%%%%%%%%%%%%%%%%%%%%%%%%%%%%%%%%%%%%%%%%%%%%%%%%%%%
\section{SU(4) symmetry in graphene}
%%%%%%%%%%%%%%%%%%%%%%%%%%%%%%%%%%%%%%%%%%%%%%%%%%%%%%%%%%%%%%%%%%%%%%%%%%%%%%
%-----------------------------------------------------------------------------
\begin{figure}[h]
\centering
\begin{minipage}{14pc}
\includegraphics[width=18pc]{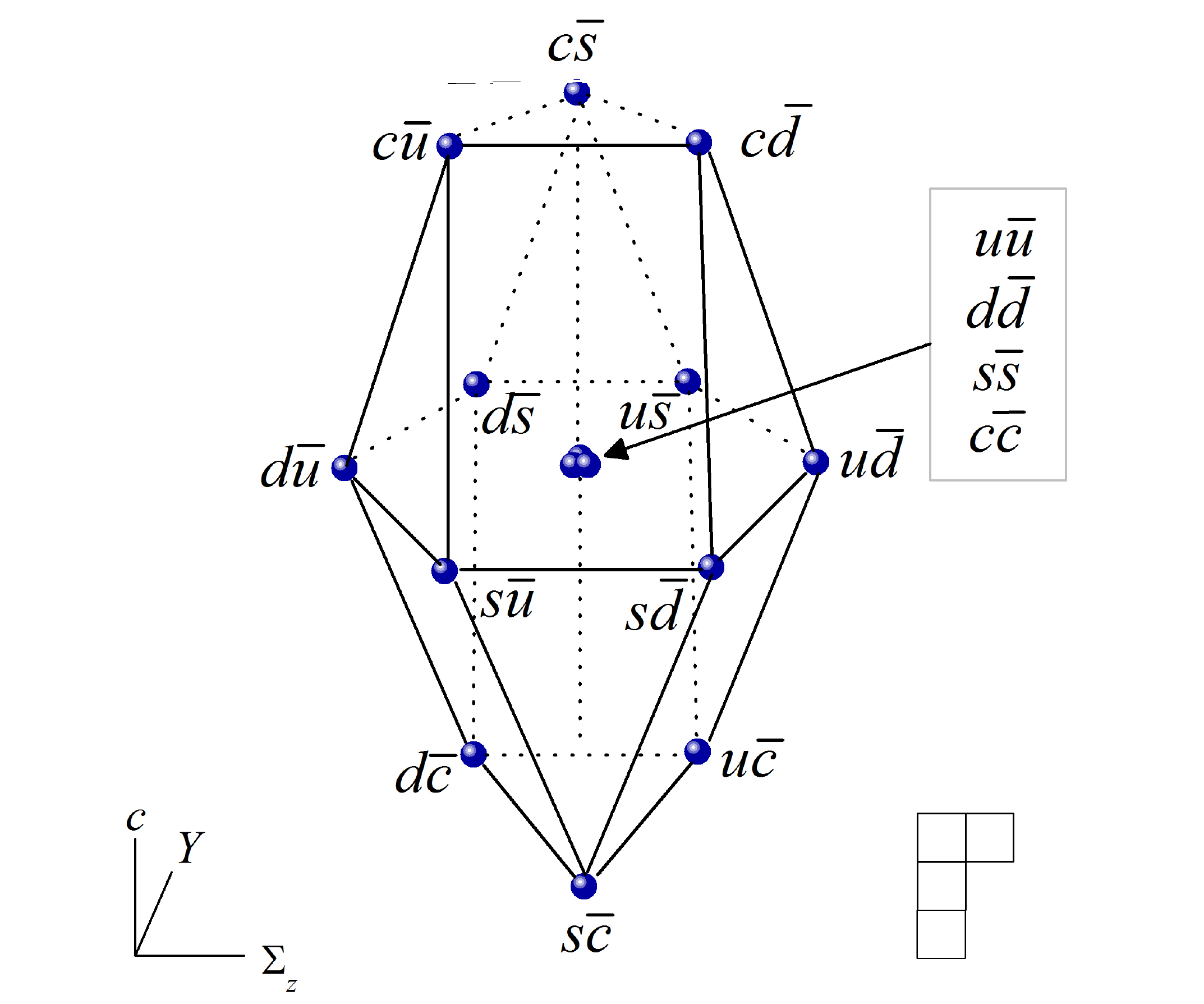}
\caption{\label{fig-mult}SU(4) $\left[15 \right]$-plet describing the excitonic
states in terms of flavour isospin ($\Sigma_z$), hypercharge ($Y$) and charm
($c$).}
\end{minipage}\hspace{4pc}%
\begin{minipage}{16pc}
\includegraphics[width=14pc]{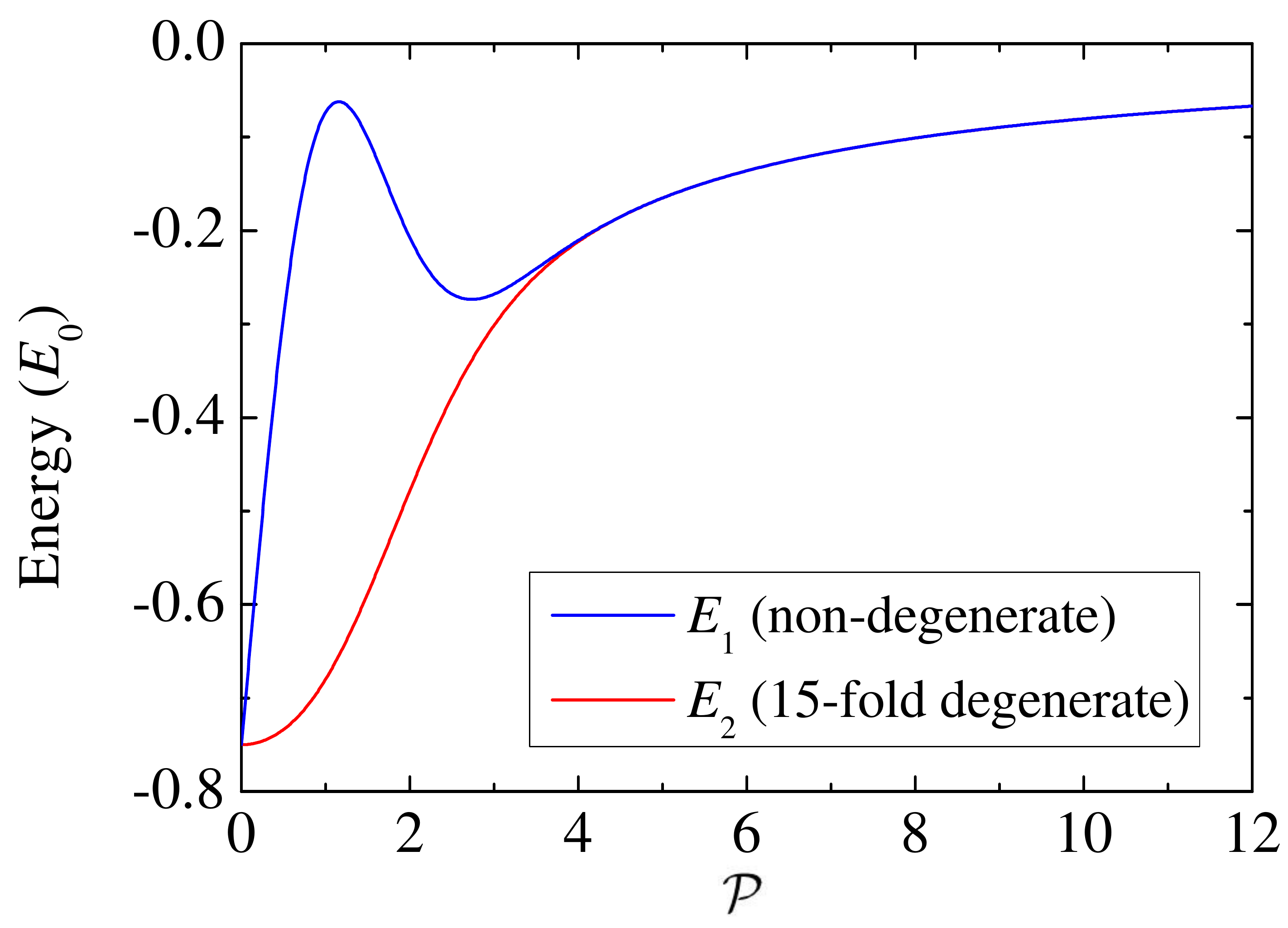}
\caption{\label{fig-disp} Dispersion relations for excitons with the hole in
$n=0$
LL and electron in $n=1$ LL. $\mathcal{P}$ is the quasimomentum in units of
$\hbar/\ell_B$. }
\end{minipage} 
\end{figure}
%-----------------------------------------------------------------------------
An electron (or hole) in graphene in a strong perpendicular magnetic field is
described by four quantum numbers: the LL index $n$; the oscillator quantum
number $m$; the spin projection $s_z$ and the pseudospin projection $\tau_z$.
We denote the fermionic creation operator for an electron by $c^{\dagger}_{nm
\tau_z s_z}$ and that for a hole by $d^{\dagger}_{nm \tau_z
s_z} \equiv c_{nm \tau_z
s_z}$. The first two orbital quantum numbers are associated with the cyclotron orbits.
%geometrical symmetries of the system. 
The last two, when considered
together, give four possible spin/pseudospin ``flavours'', leading to the
SU(4) symmetry. We identify these with the four quark flavours up ($u$), down
($d$), charm ($c$) and strange
($s$) according to $ \{ \Downarrow\downarrow, \Uparrow\downarrow,
\Downarrow\uparrow, \Uparrow\uparrow \} \equiv \{ d, u, s, c \}$. The generators
of $\mathrm{SU}(4)$ can be constructed using bilinear combinations of the
fermionic creation and annihilation operators conserving
the number of electrons and holes. 
We have
\begin{equation}
\label{eq-cij}
\hat{C}_{ij} =\sum_{nm} c^{\dagger}_{nmi} c_{nmj}
       - \sum_{nm}d^{\dagger}_{nmj} d_{nmi},
\end{equation}
where $i,j\in\left\lbrace d,u,s,c \right\rbrace$. 
The generators satisfy the
commutation relations
\begin{equation}
\label{eq-cijcomm}
[\hat{C}_{ij}, \hat{C}_{kl} ] = \delta_{jk} \hat{C}_{il} - \delta_{il}
\hat{C}_{kj},
\end{equation}
so are closed under commutation, confirming that they form a Lie algebra. The operators associated with spin and pseudospin may of course be expressed in terms of the generators. For example, $S_z=\frac{\hbar}{2}\left(\hat{C}_{ss}+\hat{C}_{cc}-\hat{C}_{uu}-\hat{C}_{dd}\right)$ and $S_+=S_x+iS_y=\hbar\left(\hat{C}_{cu}+\hat{C}_{sd}\right)$.

In particle physics, mesons, which are a bound quark-antiquark pair, are
classified using Young diagrams \cite{GreM97}. We use the same techniques to
determine the
multiplet structure for electron-hole ($e$-$h$) complexes in graphene. Young
diagrams describe SU($n$) multiplets and consist of $n$ or fewer left-justified
rows of boxes, such that each row is as least as long as the row beneath it.
Roughly speaking, each box represents a particle. Each $\mathrm{SU}(n)$
multiplet is uniquely labelled by a list of $n-1$ non-negative integers,
$(a_1,a_2,\ldots,a_{n-1})$. The integer in the $i^{\mathrm{th}}$ position
corresponds to
the number of boxes in the $i^{\mathrm{th}}$ row minus the number of boxes in
the $(i+1)^{\mathrm{th}}$ row i.e. the overhang of the $i^{\mathrm{th}}$ row. By
the definition
of a multiplet label, adding columns of length $n$ to the left of a Young
diagram is
redundant. Young diagrams can be used to calculate the multiplicity, $[N(n)]$,
of an SU(n) multiplet. For the $n=4$ case of current interest,
\begin{equation}
\label{eq-nof4}
[N(4)]=\frac{(a_1+1)}{1}\frac{(a_2+1)}{1}\frac{(a_3+1)}{1}\frac{(a_1+a_2+2)}{2}
\frac{(a_2+a_3+2)}{2}\frac{(a_1+a_2+a_3+3)}{3}.
\end{equation}
There are two fundamental representations of $\mathrm{SU}(n)$: $[n]$, with Young
diagram $\yng(1)$ and label $(1,0,\ldots0)$ and $[\bar{n}]$ with Young diagram
composed of $n-1$ vertically stacked boxes and label $(0\ldots,0,1)$. The latter
represents an antiparticle; this makes sense if we identify $n$ vertically
stacked boxes with the vacuum. We shall also use the symbol, $\underline{1}$ to
denote the vacuum. For the current purpose, the hole in a MP behaves as an
antiparticle. Thus the electron has Young diagram $\yng(1)$ and the hole Young
diagram $\yng(1,1,1)$. To determine the SU(4) multiplet structure for the MPs,
we must combine these two diagrams. The rules for combining Young diagrams
\cite{GreM97} yield
\begin{equation}
\label{eq-youngmultsmeson}
\yng(1) \otimes \yng(1,1,1) = \yng(2,1,1) \oplus \underline{1}.
\end{equation}
Eq.\ (\ref{eq-youngmultsmeson}) is equivalent to $(1,0,0) \otimes (0,0,1) =
(1,0,1) \oplus (0,0,0)$ or $[4]\otimes[\bar{4}]=[15]\oplus[1]$. Now the
generators, $\hat{C}_{ij}$, commute with the interacting Hamiltonian describing
the electron system up to some small symmetry breaking terms \cite{GoeMD06}. In
addition,
the application of a generator to a state in a particular multiplet always
returns a state in that same multiplet with the same Coulomb interaction energy. This ensures that states within a given
multiplet are degenerate in their energies. Thus from this simple
calculation, we have shown that the MPs in graphene will either have degeneracy
1 or 15. The [15]-plet is shown in Fig.\ \ref{fig-mult} and is analogous to the meson multiplet of particle physics. The states are plotted
as a function of:
the flavour isospin projection $\hat{\Sigma}_z = \frac{1}{2}\left( \hat{C}_{uu}
- \hat{C}_{dd}\right) $,
the hypercharge $\hat{Y} =
\frac{1}{3}\left(\hat{C}_{uu}+\hat{C}_{dd}+\hat{C}_{cc}
\right)-\frac{2}{3}\hat{C}_{ss} $
and the charm $\hat{c} = \hat{C}_{cc}$.
One may deduce the symmetry of states under exchange of
quantum numbers for different particles from their Young diagrams \cite{GreM97}.
%%%%%%%%%%%%%%%%%%%%%%%%%%%%%%%%%%%%%%%%%%%%%%%%%%%%%%%%%%%%%%%%%%%%%%%%%%%%%%
\section{Dispersion relation for magnetoplasmons in pristine
graphene}
%%%%%%%%%%%%%%%%%%%%%%%%%%%%%%%%%%%%%%%%%%%%%%%%%%%%%%%%%%%%%%%%%%%%%%%%%%%%%%
As a precursor to our results for localised MPs in dirty graphene, we
discuss the dispersion relation for extended MPs in a clean system. The
calculation is explained in Ref.\ \cite{IyeWFB07} in the same spirit as that
for the 2DEG established by Kallin and Halperin \cite{KalH84}, but with the
additional complication of the valley pseudospin degree of freedom. The centre
of mass quasimomentum for the exciton, $\mathbf{P}$, can be shown to be a good
quantum number. One can think of this as being due to the repulsive Lorentz
forces and attractive Coulomb forces, which act on the electron and hole,
cancelling out when they have a certain separation, enabling the exciton's
centre of mass to move in a straight line. 
\begin{figure}
 \centering
 \includegraphics[width=0.45\textwidth]{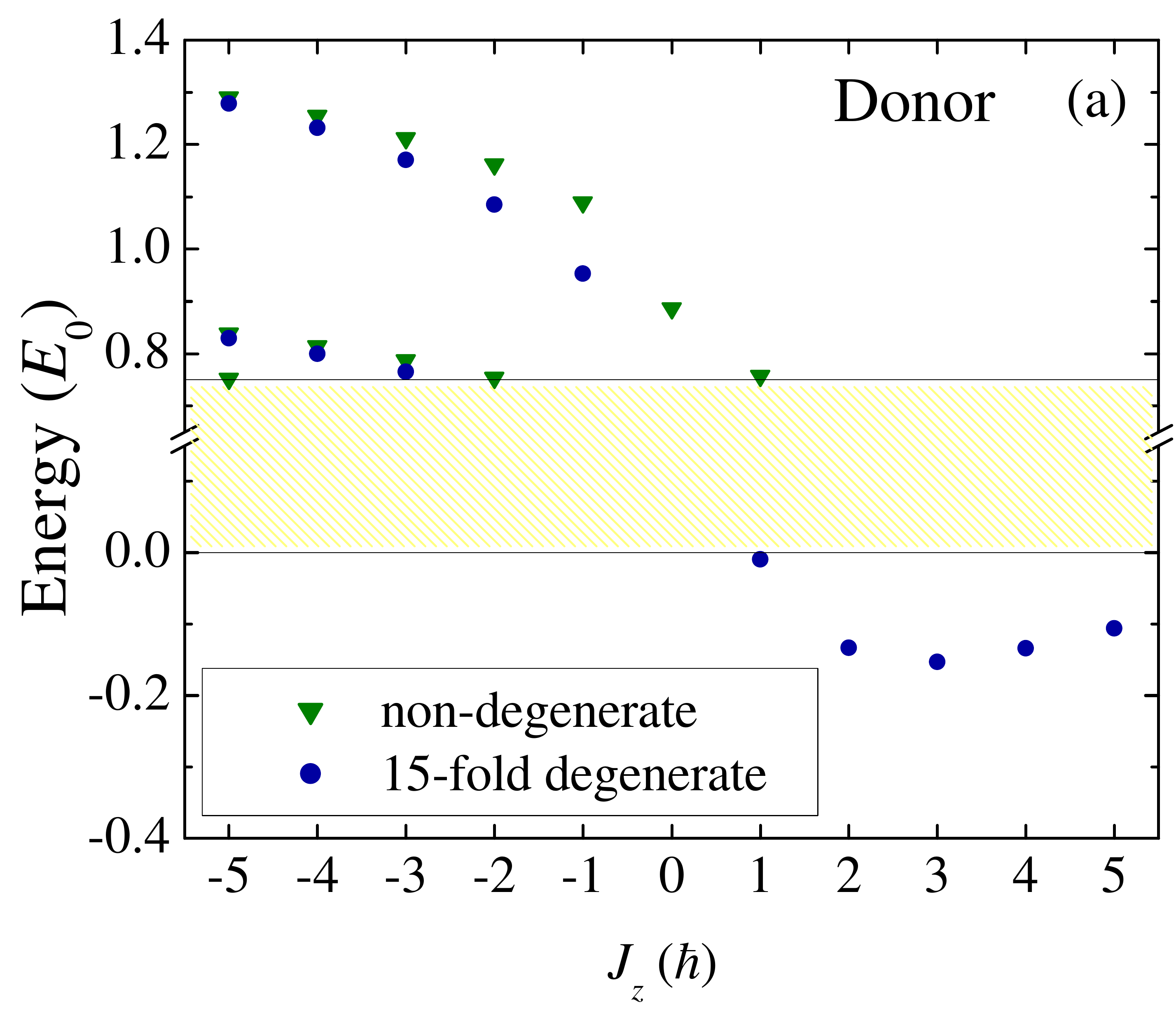}
 \includegraphics[width=0.45\textwidth]{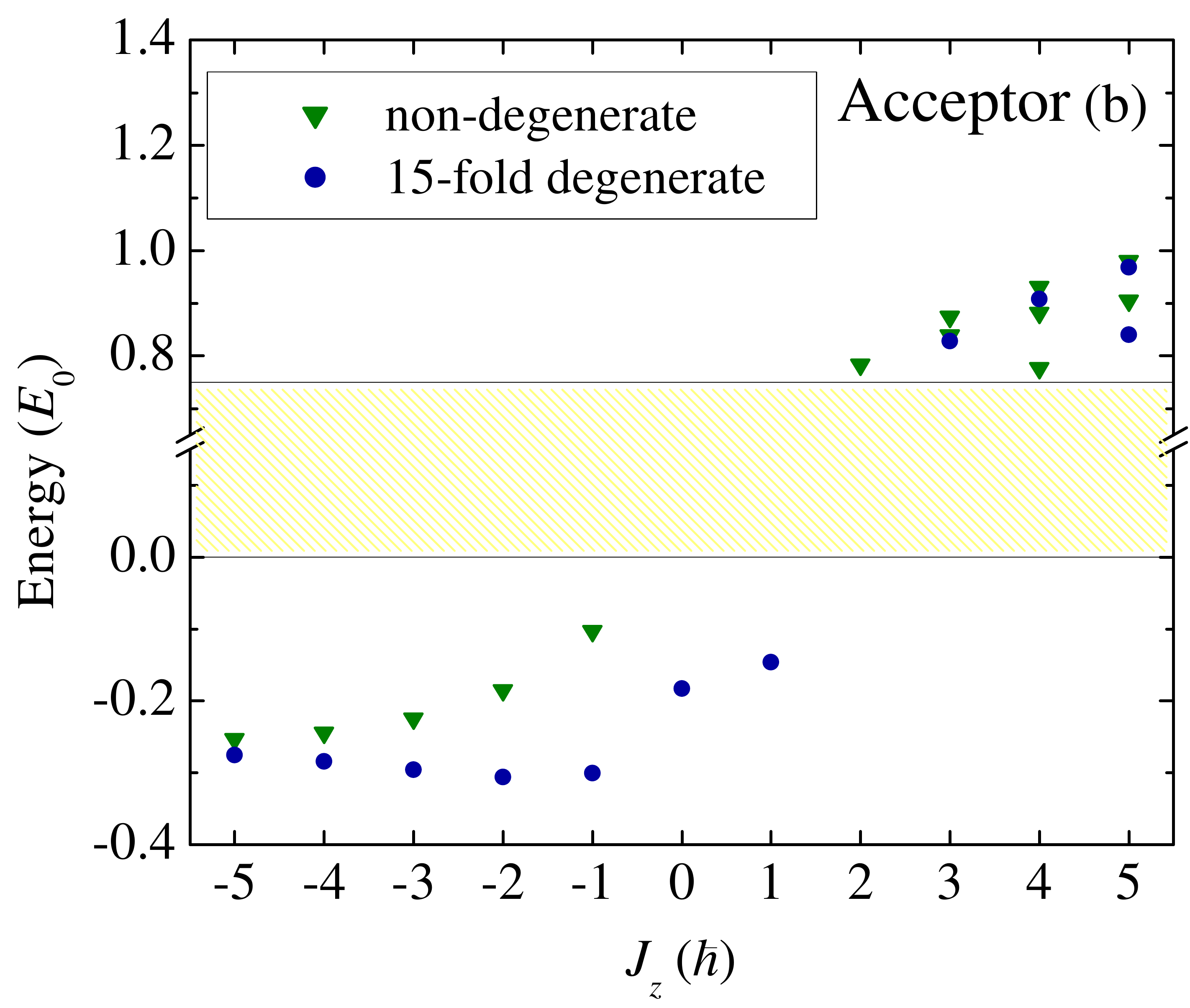}
  \caption {MPs from the ground state $\nu=2$ localised on (a) a donor impurity
and (b) an acceptor impurity. The hatched area of width $0.75 E_0$ represents
the continuum of extended MPs. Energies are given relative to the lower
continuum edge in units of $E_0$ as a function of the generalised angular
momentum projection, $J_z$. 
}
\label{fig-emz}
\end{figure}
The Hamiltonian (in the symmetric
gauge) for MPs is \cite{FisDR09a}:
\begin{eqnarray}
\label{eq-ham} %(1)
\hat{H} & = & \sum_{\mathcal{N},m} \tilde{\epsilon}_n c^{\dag}_{\mathcal{N}
m}c_{\mathcal{N} m}-\sum_{\mathcal{N},m} \tilde{\epsilon}_n
d^{\dag}_{\mathcal{N} m}d_{\mathcal{N} m} \nonumber\\ 
& + & \sum_{\substack{\mathcal{N}_1, \mathcal{N}_2 \\ \mathcal{N}_1',
\mathcal{N}_2'} }\sum_{\substack{m_1, m_2 \\ m_1', m_2'}
}\mathcal{W}_{\mathcal{N}_1 m_1   \mathcal{N}_2 m_2}^{\mathcal{N}_1' m_1' 
\mathcal{N}_2' m_2'}c^{\dag}_{\mathcal{N}_1' m_1'}d^{\dag}_{\mathcal{N}_2'
m_2'}d_{\mathcal{N}_2 m_2}c_{\mathcal{N}_1 m_1},
\end{eqnarray}
where we define the collective index $\mathcal{N}=\{n \tau_z s_z \}$.
The first two terms give the single particle contribution. The
$\tilde{\epsilon}_n$ are the LL energies renormalised by a self energy
correction due to exchange interactions. The last term describes the
$e$-$h$ interaction; it has direct and exchange components. This
Hamiltonian mixes the four excitons for which there are no spin or pseudospin
flips and leaves the remaining twelve unmixed. 
The Coulomb interaction is treated exactly in the span  of given LLs and the dispersion relations are found by diagonalising the Coulomb interaction Hamiltonian in the basis of two-particle electron-hole states compatible with magnetic translations (having definite quasimomentum P).
% To determine the dispersion
% relation, the non-interacting $e$-$h$ wavefunction, which depends on quasimomentum,
% is first found by manipulating the single particle Hamiltonian through
% coordinate transformations and a gauge transformation until it can be formulated
% in terms of Bose ladder operators. This is used to find the secondary quantised
% operator, which creates such an $e$-$h$ pair out of the Dirac sea. 
The Coulomb
interaction is treated as a perturbation and the dispersion relation found by
diagonalising the Coulomb interaction matrix in the basis of these
non-interacting two-particle wavefunctions. For the fully filled LL
considered here, there are only two possible dispersion relations. The four
mixed excitons give a non-degenerate dispersion branch \cite{IyeWFB07}
\begin{equation}
E_1=-\frac{E_0}{8}e^{-\frac{\mathcal{P}^2 }{4 }}\left[\left(6 +  \mathcal{P}^2    \right)I_0\left(\frac{\mathcal{P}^2}{4} \right)- \mathcal{P}^2 I_1\left(\frac{\mathcal{P}^2}{4} \right)\right]
\label{eq-e1}
\end{equation}
and a triply degenerate dispersion branch 
\begin{equation}
E_2=E_1+E_0\mathcal{P}e^{-\frac{\mathcal{P}^2 }{2}},
\label{eq-e2}
\end{equation}
where $\mathcal{P}=P\ell_B/\hbar$ is the dimensionless centre of mass quasimomentum, $P$, for the exciton and $I_0$ and $I_1$ are modified Bessel functions. The dispersion, $E_2$, is the same as for tthe 12
unmixed excitons, making the total degeneracy 15, as predicted in the previous
section. These dispersion relations are plotted in Fig.\ \ref{fig-disp}.
Energies are given in units of $E_0=\sqrt{\frac{\pi}{2}}\frac{e^2}{\varepsilon
\ell_B}$, the characteristic Coulomb interaction energy, where $\varepsilon$ is
an effective dielectric constant.

%%%%%%%%%%%%%%%%%%%%%%%%%%%%%%%%%%%%%%%%%%%%%%%%%%%%%%%%%%%%%%%%%%%%%%%%%%%%%%
\section{Localised magnetoplasmons in dirty graphene}
%%%%%%%%%%%%%%%%%%%%%%%%%%%%%%%%%%%%%%%%%%%%%%%%%%%%%%%%%%%%%%%%%%%%%%%%%%%%%%
We now consider how MPs may become localised in the presence of a single
Coulomb impurity with potential, $V(r)=\frac{Ze^2}{\varepsilon r}$, where
%$\varepsilon$ is the effective dielectric constant allowing for some screening
$Z$ is the impurity charge in units of $e$ \cite{FisDR09a}. The centre of mass
quasimomentum is
no longer well defined for a MP. However, the axial symmetry remains and the
orbital angular momentum projection, defined in terms of the electron and hole
orbital quantum numbers as, $J_z=|n_e|-|n_h|-m_e+m_h$, is a good quantum number.
In the
presence of an impurity, localised MPs break away from the continuum of
extended MPs as discrete states. These localised MPs are shown for a donor
($Z=1$) and acceptor ($Z=-1$) impurity in Fig. \ref{fig-emz}, plotted as a
function of $J_z$. The yellow shaded area represents the continuum; it has width
$0.75E_0$, as can be seen in Fig.\ \ref{fig-disp}. Note that for both impurity
cases, there are non-degenerate branches and branches with 15-fold degeneracy. 
The predictions based on SU(4) symmetry hold here, because the Coulomb
impurity, being long ranged, does not introduce an inequivalency between the
valleys. This would not be the case for a short-ranged impurity located on one
of the carbon sites \cite{FisDR09b}.
% \begin{figure}[h]
% \begin{minipage}{14pc}
% \includegraphics[width=14pc]{fig-emz-don}
% \caption{\label{label}Figure caption for first of two sided figures.}
% \end{minipage}\hspace{2pc}%
% \begin{minipage}{14pc}
% \includegraphics[width=14pc]{fig-emz-accptr}
% \caption{\label{label}Figure caption for second of two sided figures.}
% \end{minipage} 
% \end{figure}
%%%%%%%%%%%%%%%%%%%%%%%%%%%%%%%%%%%%%%%%%%%%%%%%%%%%%%%%%%%%%%%%%%%%%%%%%%%%%%
\section{Conclusions}
%%%%%%%%%%%%%%%%%%%%%%%%%%%%%%%%%%%%%%%%%%%%%%%%%%%%%%%%%%%%%%%%%%%%%%%%%%%%%%
In conclusion, we have studied neutral collective excitations of graphene in
the presence of a strong perpendicular magnetic field. Electrons and holes in
graphene can exist in four possible spin/pseudospin states, so that
electron-hole complexes may be likened to mesons composed of quarks with four
possible flavours. This analogy enabled us to use techniques established for
treating mesons to obtain the multiplet structure of the neutral
collective excitations for
both extended excitations and those localised on a Coulomb impurity. An
extension of this idea to charged collective excitations, which can be
thought of as three particle complexes, can be found in Ref.\ \cite{FisRD10}.
These are analogous
to baryons formed of three quarks.
\ack We acknowledge funding by EPSRC (͑AMF and RAR). AMF is
grateful for hospitality at CSU Bakersfield. ABD acknowledges Cottrell Research
Corporation and the Scholarship of
KITP, UC Santa Barbara.

\end{document}